\pdfoutput=1
\documentclass[twocolumn,prl,preprintnumbers,amsmath,amssymb,floatfix]{revtex4}
\usepackage[pdftex]{graphicx}
\usepackage{color}
\usepackage{dcolumn}
\usepackage{bm}
\usepackage{subfigure}
\usepackage{times}
\usepackage{amsmath}
\usepackage{amssymb}

\begin{document}

\title{Scaling behavior in convection-driven Brazil-nut effect}

\author{Prakhyat Hejmady}
\email{phejmady@rri.res.in}
\author{Ranjini Bandyopadhyay}
\email{ranjini@rri.res.in}
\author{Sanjib Sabhapandit}
\email{sanjib@rri.res.in}
\author{Abhishek Dhar}
\email{dabhi@rri.res.in}
\affiliation{Raman Research Institute, CV Raman Avenue, Sadashivanagar, Bangalore 560080, India}

\vspace{0.5cm}

\date{\today}

\begin{abstract} 
The Brazil-nut effect is the phenomenon in which a large intruder
  particle immersed in a vertically shaken bed of smaller particles
  rises to the top, even when it is much
  denser.  The usual
  practice, while describing these experiments, has been to use the
  dimensionless acceleration $\Gamma=a \omega^2/g$, where $a$ and
  $\omega$ are respectively the amplitude and the angular frequency of
  vibration and $g$ is the acceleration due to
  gravity.
  Considering a vibrated quasi-two-dimensional bed of mustard seeds,
  we show here that the peak-to-peak velocity of shaking $v= a\omega$,
  rather than $\Gamma$, is the relevant parameter in the regime where
  boundary-driven granular convection is the main driving mechanism.
  We find that the rise-time $\tau$ of an intruder is described by the
  scaling law $\tau \sim (v-v_c)^{-\alpha}$, where $v_{c}$ is identified as
  the critical vibration velocity for the onset of convective motion
  of the mustard seeds. This scaling form holds over a wide range of
  $(a,\omega)$, diameter and density of the intruder.
\end{abstract}

\maketitle

Besides its obvious technological importance, granular matter is one
of the most interesting examples of a driven dissipative system.
Shaken granular matter exhibits various phenomena such as
convection, segregation and jamming. 
A striking experimental observation in vibrated granular
  materials is the Brazil-nut effect (BNE), where a large intruder
  particle immersed in a vertically shaken bed of smaller particles
  rises to the top, even when it is much
  denser~\cite{rosato87,JNB96,mobius01,kudrolli04,duran94,knight96,vanel97,liffman01}.  
A related
effect is the segregation obtained on vibrating a mixture of particles
of different sizes and densities.  While some progress has been made,
a complete understanding of these observations is still lacking.  The
consensus now is that different segregation mechanisms such as
convection, void filling, air-drag and kinetic-energy driven methods
are responsible for the effect, each being dominant in different
parameter regimes \cite{JNB96,mobius01,kudrolli04,cooke96,schroter06}.
In this Letter, we show that in a broad parameter regime where convection
is responsible for BNE, the  driving velocity  $v = a \omega$ is the relevant scaling variable  rather than the commonly used dimensionless acceleration $\Gamma=a \omega^2/g$. Here, $a$ 
and $\omega$ are respectively the amplitude and angular frequency of vibration and $g$ is the acceleration due to gravity. This is demonstrated through a   
very good  collapse of the rise-time data  
 as a function of the driving velocity.

It is well known that
granular convection can be either boundary wall-driven~\cite{knight96} or
buoyancy-driven~\cite{eshuis07}. Boundary-driven convection is unique to granular
systems and refers to the case where the surface properties of the
side walls and their shape are crucial in determining the formation
and nature of the convection
rolls~\cite{knight93,ehrich95,BM95,knight97}. Buoyancy-driven granular
convection, on the other hand, is seen under strong driving and is analogous 
to Rayleigh-Benard  convection in fluids~\cite{eshuis05,eshuis07,alam10}.

\begin{figure}
\begin{center}
\includegraphics[width=3.3in]{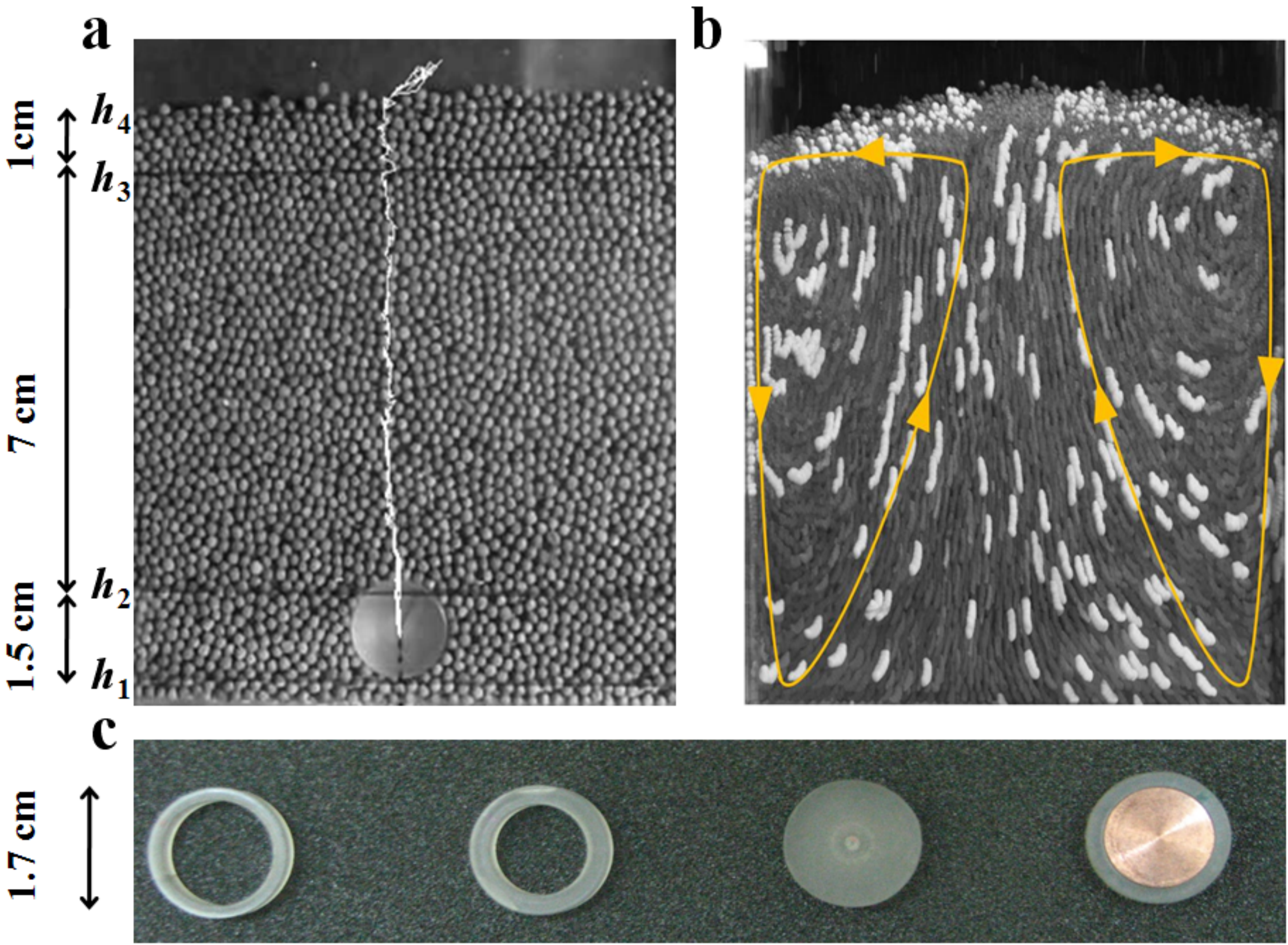}
\caption{ (Color online) {Experimental setup}. {\bf a}. The initial position of
  an acrylic intruder of diameter $1.7$~cm embedded in a column of
  mustard seeds of sizes between $1.5$~mm and $2$~mm. The bottom of the 
intruder is placed at a height $h_1=5$~mm from the floor of the cell. The rise-time
  $\tau$ is the time taken by the centre of the intruder to move from
  the height $h_{2}$ to the height $h_{3}$.  A typical trajectory of
  the rising intruder is shown by the vertical white line. {\bf
    b}. Convection pattern obtained by superposing many consecutive
  time-lapsed images of the vibrated bed of black mustard seeds in which yellow 
mustard seeds are 
distributed randomly to serve as tracer particles. The two loops (yellow)
  with arrows indicate the flow direction of the mustard seeds. 
{\bf c}. Intruders used in
  the experiment with relative densities ranging between $0.30$ --
  $2.34$.}
\label{setup}
\end{center}
\end{figure}
\begin{figure}
\begin{center}
\includegraphics[width=\hsize]{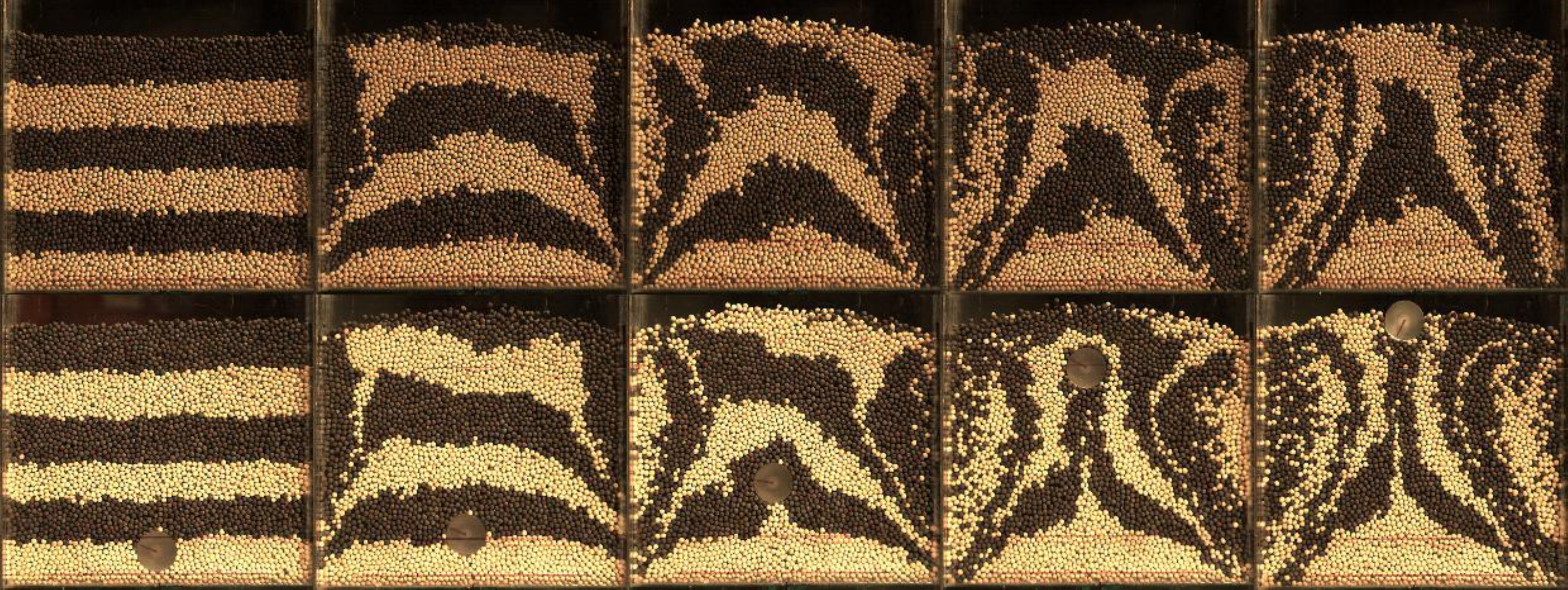}
\caption{ (Color online) {A series of snapshots of the vibrated bed taken at
    regular intervals of $8$ seconds }. Top and bottom panels are for
  experiments done without and with an intruder respectively. The
  shaking parameters were set at $f=50$~ Hz and  $a=1.6$~mm. In the bottom panel, the intruder diameter is $1.7$~cm  and
  $\rho_r=1.07$.  The alternate black and yellow mustard layers
  initially destabilize at the walls and eventually bulk convection
  rolls are formed.  An inspection of these figures confirms that the granular
  convection patterns do not change significantly as a result of the
  presence of the intruder, except close to the intruder boundary.  }
\label{snaps}
\end{center}
\end{figure}
  In this Letter, our focus is on boundary-driven convection.
One of the simplest measurements in a granular convecting fluid is
that of the rise-time $\tau$ of a large intruder. This quantity
effectively provides a single observable characterizing the complex
velocity field in the system. The intruder rise-time has been used as
a probe for understanding the effects of driving frequency and
amplitude, interstitial air-flow, the size and density of the
intruder, and wall effects on BNE.  Duran {\it et al.} \cite{duran94}
studied BNE in a quasi-two-dimensional bed of aluminium beads. They
demonstrated two different segregation mechanisms: arching effects were found to dominate in
the range $ \Gamma_c < \Gamma \lesssim 1.5$ with $\Gamma_c \approx 1$,
while convection gives rise to segregation at $ \Gamma \gtrsim 1.5$. In the arching
regime, intruder rise was observed only when intruder-to-bead diameter
ratio was large enough. In the convective regime, the rise-time was
independent of intruder diameter.  For a three-dimensional bed of
glass beads, the rise-time
of tracer particles in the boundary-driven convection regime was studied in 
\cite{knight96}. For a
\emph{fixed} driving frequency, a $\tau \sim
(\Gamma-\Gamma_c)^{-2.5}$ dependence with $\Gamma_c \approx 1.2$ was reported,
while for fixed $\Gamma$ values, an exponential
dependence on frequency was obtained. In a later work, Vanel {\it et al.}
\cite{vanel97} looked at the motion of an intruder and reported
deviations from the above behaviour. However, none of these studies
have obtained any scaling of the rise-time data. An understanding of
the dependence of rise-time on driving amplitude and frequency
therefore remains incomplete.

The present work reports experimental measurements of the rise-time of
an intruder in a vertically vibrated quasi-two-dimensional bed of
mustard seeds (Fig.~\ref{setup}).  The experimental setup  consists of a
quasi-two-dimensional rectangular acrylic cell of height $25$ cm,
width $13$ cm and gap $6$ mm.  The cell is filled up to a height $h$
with black mustard seeds of density $1.11$ g/cc and sizes between
$1.5$ and $2$ mm. The intruder is a large acrylic disk of diameter
$1.7$ cm, thickness $5$ mm and density $1.07$ g/cc. The thickness of
the intruder is chosen such that mustard seeds are not able to enter
the space between the intruder and the cell walls. The cell is mounted
on a $50$ kgf DESPL electrodynamic shaker executing vertical
sinusoidal vibrations of the required amplitudes and frequencies.  The
narrow side walls of the cell are roughened by lining them with fine
sandpaper (ISO grit designation $P-100$). For these boundary
conditions, the experiments were reproducible. However, for smooth
walls it was found that the convection patterns were irreproducible
and often the system would jam.  The humidity of the lab was
maintained at $40\%$.

\begin{figure}
\begin{center}
\includegraphics[width=3.3in]{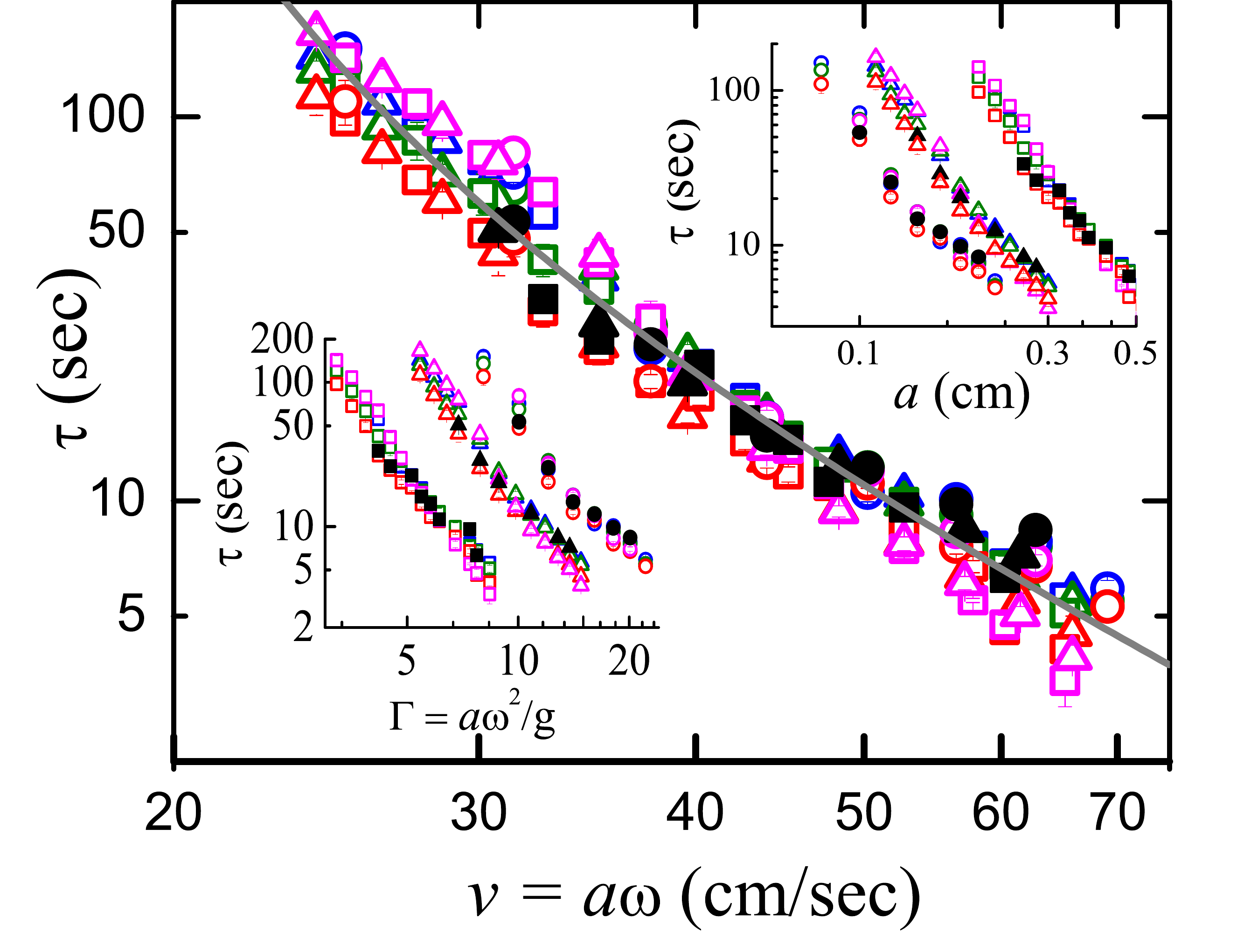}
\caption{ (Color online) { Intruder rise-time plots}. {Upper inset}: Plots of
  the rise-time $\tau$ versus shaking amplitude $a$ for different
  frequencies: i) $20$ Hz ($\square$) ii) $35$ Hz ($\triangle$) and
  iii) $50$ Hz ($\bigcirc$) for an intruder of diameter $1.7$ cm and
  for four different relative densities $\rho_{r}$ = 0.3 (blue), 0.8
  (green), 1.07 (red) and 2.34 (magenta). The filled symbols
  correspond to a $1$~cm diameter intruder with $\rho_r=1.07$. {
    Lower inset}: Plot of the same data versus the dimensionless
  acceleration $\Gamma$.  { Main figure}: Plot of $\tau$ versus
  peak-to-peak velocity of shaking $v = a\omega$.  The data collapses
  for the entire range of shaking velocities, for all relative
  densities and for the two intruder sizes.  The solid line shows the
  functional form $\tau \sim (v-v_{c})^{-\alpha}$ with $\alpha=2.0$ and $v_c=16$ cm/sec.}
\label{risetime}
\end{center}
\end{figure}

In order to produce fixed initial conditions corresponding roughly to
a fixed spatial distribution of mustard seeds, the cell was initially
vibrated at a high $\Gamma$ and the intruder was pushed to a fixed
depth $h_1$ (Fig.~\ref{setup}a). The driving was then removed and the
top layer leveled.  The cell was then vibrated at the required amplitude  and
frequency ($f=\omega/2\pi$).  The upward motion of the intruder was recorded using a Panasonic
Lumix DMC-TZ3 digital camera at a rate of $33$ frames/sec. 
 The tracking of the
intruder (Fig.~\ref{setup}a) 
was done using a Labview based particle tracking algorithm
\cite{StAndrews}.  
In order to change the density of the intruder,
several acrylic intruders with holes in the centre were made. By changing
the size of the holes or inlaying them with copper, a range of
density ratios $\rho_r =({\rm intruder~density})/({\rm
  mustard~seed~density})$ between $ 0.30$ and $2.34$ (Fig.~\ref{setup}c) was
achieved: $\rho_r= 0.3$ for acrylic with $1.4$ cm air hole, $0.8$ for
acrylic with $1.2$ cm air hole, $1.07$ for solid acrylic disk and
$2.34$ for acrylic with $0.65$ cm copper.
The rise-time of the intruder was computed by estimating the time
taken by it to rise from a fixed starting position $h_{2}$
to a fixed final position $h_{3}$ (Fig.~\ref{setup}a). The mean
rise-time $\tau$ was computed by averaging over $10$ trials for every
parameter set.

\begin{figure}
\begin{center}
\includegraphics[width=3.3in]{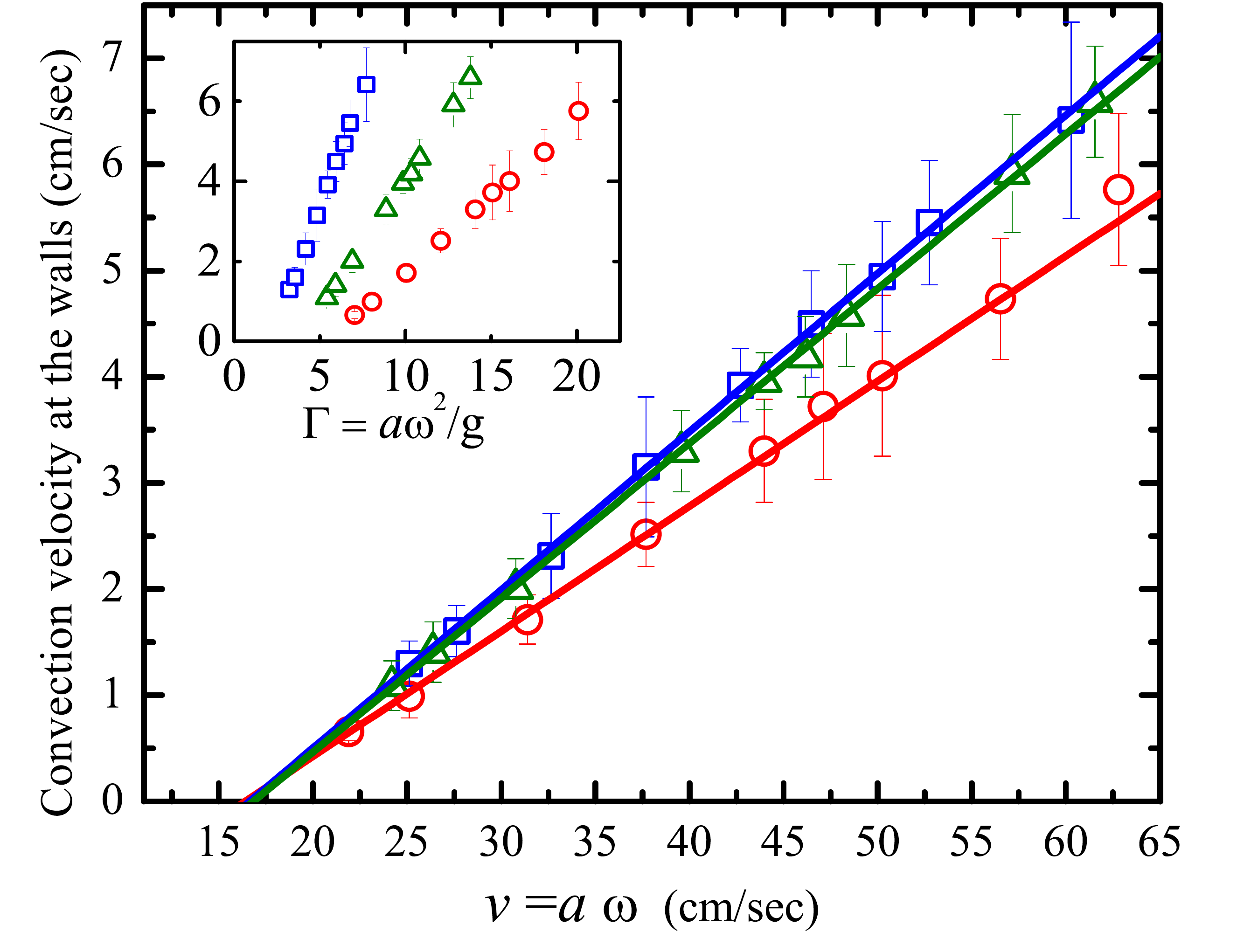}
\caption{(Color online) {Convection velocities at the side walls}. The convection
  velocities, in the absence of the intruder, are plotted versus the
  peak-to-peak velocity of shaking $v = a\omega$ for shaking
  frequencies: i) $20$ Hz (${\color{blue}\square}$), ii) $35$ Hz
  (${\color{green} \triangle}$) and iii) $50$ Hz
  (${\color{red}\bigcirc}$). Each linear fit (denoted by solid line)
  is extrapolated to estimate the peak-to-peak shaking velocity for
  the onset of convection. The $x$-intercepts lie in the range of $16$
  -- $17$ cm/sec. {Inset}: Plot of the convection velocities at
  the side-wall as a function of $\Gamma$. }
\label{velM}
\end{center}
\end{figure}

 The measurements were made under a
fairly wide range of driving parameters, obtained by varying
frequencies and amplitudes ($2.5 \lesssim \Gamma \lesssim 25$),
intruder densities and diameters. 
 Shaking the bed vertically results in a thin
downward moving layer of mustard seeds adjacent to each side wall,
which sets up bulk upward motion elsewhere in the system as seen in 
Fig.~\ref{setup}(b). This convective motion of the medium carries the
intruder to the top of the bed.  The granular convection phenomenon was studied in greater detail by
filling up the rectangular cell with alternating layers of black and
yellow mustard seeds, with each layer being approximately $2$ cm thick. In 
Fig.~\ref{snaps} we see that for  a given set of shaking parameters, 
very similar flow patterns of the medium are obtained with and without the
intruder

Next, we measure the rise-times $\tau$ for several
peak-to-peak amplitudes and frequencies and for
various relative densities $\rho_r$.  The rise-time decreases with increasing
shaking frequencies and amplitudes, as seen in the upper inset of
Fig.~\ref{risetime}.  It is observed that $\tau$ is almost independent
of density over the range used in these experiments ($\rho_r=0.30$ --
$2.34$). This is in contrast to the three-dimensional
results \cite{mobius01,mobius04,mobius05} where a non-monotonic
dependence of $\tau$ on $\rho_{r}$ was found.  There is no scaling of
$\tau$ with $\Gamma$ (lower inset of Fig.~\ref{risetime}).  In
contrast, a plot of $\tau$ as a function of the shaking velocity $v=a
\omega$ (Fig.~\ref{risetime}) shows an excellent collapse of the data.
Throughout the parameter regime where the collapse holds, we find that
boundary driven convection is the dominant mechanism responsible for
the rise of the intruder. This collapse holds for the entire range of intruder 
densities and the
two intruder sizes (diameters $1.7$~cm and $1$~cm) that were used. The intruder 
therefore behaves
approximately like a massless tracer particle tracking the convective
flow of the background granular medium. The
collapsed $\tau$-vs-$v$ data of Fig.~\ref{risetime} is well described
by the scaling-law $\tau = A (v-v_{c})^{-\alpha}$, where $\alpha=2.0$ and  $A$, $v_{c}$ are 
fitting parameters. 
For the range of $\rho_{r}$ values investigated,
$v_{c}$ is found to be around $15$ -- $18$ cm/sec and $A$ is around $13,000 \pm 500$ cm$^2$/sec.  We note that $v$ has been used as a
scaling parameter in the vibro-fluidized
regime~\cite{eshuis05,eshuis07,alam10} that is typically realized at
much higher driving ($\Gamma \gtrsim 50$) and is \emph{not}
wall-driven. It has also been noted in experiments on granular compaction by tapping that $v$,  rather than $\Gamma$,  is the appropriate  control parameter \cite{dijksman09}.
However, no scaling has been achieved so far in the
boundary-driven convection regime.  Given the non-homogeneous flow
fields that are set up in our experiments, the dependence of $\tau$ on
shaking velocity is rather surprising.

The absence of any dependence of the rise-time of the intruder on its size and density implies that convection is the dominant driving
mechanism.  To confirm this premise, we make an independent
measurement of the convection velocity of the mustard seeds without
the intruder. For measuring the convection velocity at the wall, tagged particles (yellow 
mustard seeds of sizes and densities lying within the same range as the black mustard seeds)  
were  distributed randomly in the bed of black mustard seeds. 
Over a range of shaking frequencies and amplitudes, we
measured the velocity of these tagged  particles moving down the 
two side walls (supplementary material).  
High speed
tracking of mustard-seed convection at the side walls was done using
an IDT MotionPro high speed camera at a rate of $150$ frames/sec. 
The velocity of the boundary layer
particles is an order of magnitude larger than the mean velocity of
the intruder.  As shown in Fig.~\ref{velM}, the convection velocities measured by this procedure  show a much better data-collapse when plotted against $v=a \omega$ than when plotted as a 
function of $\Gamma$.
  A linear extrapolation of the data to zero
convection velocity gives a critical velocity of $16$ -- $17$ cm/sec
for the onset of convection.  This critical velocity range overlaps
with the range of the fitting parameter $v_c$ obtained from the
rise-time data.  It is therefore natural to identify $v_c$ as the
critical shaking velocity required for the onset of convection cycles
in the granular column.

\begin{figure}
\begin{center}
\includegraphics[width=3.3in]{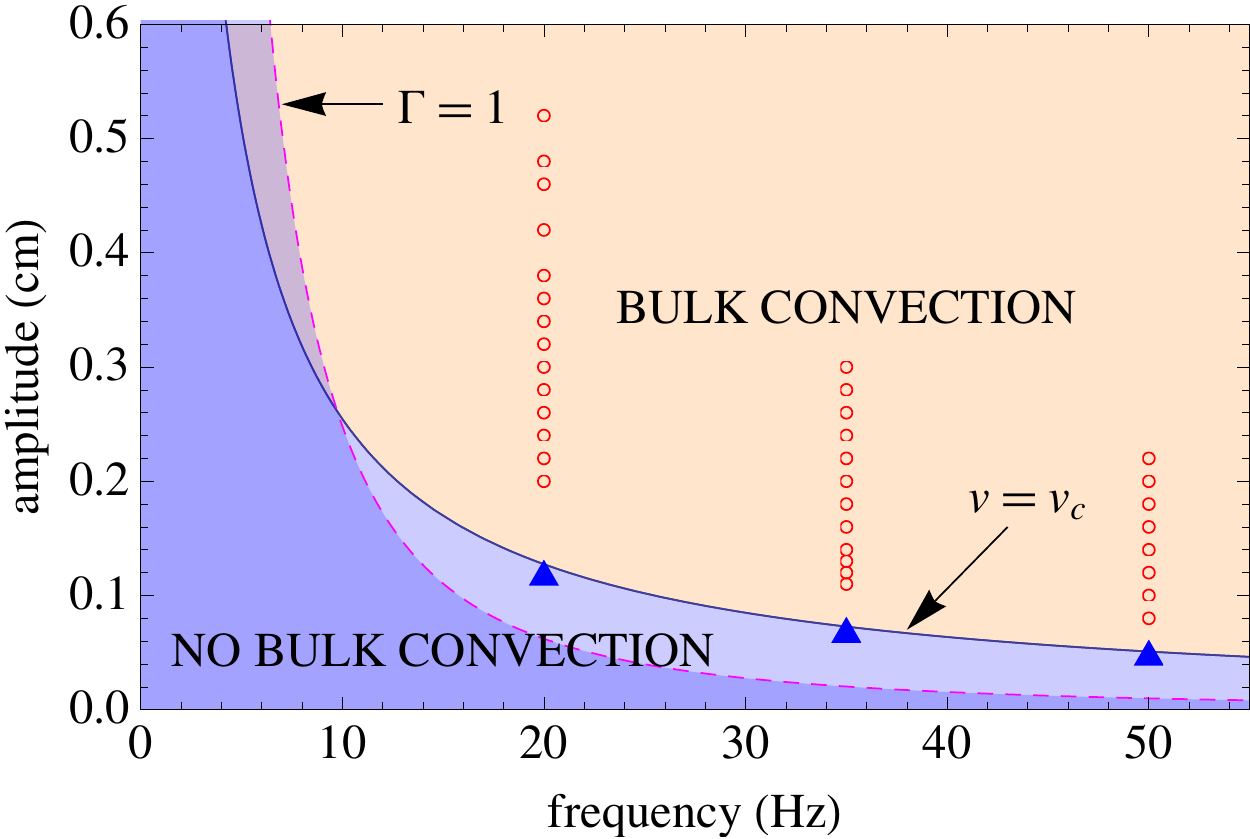}
\caption{(Color online) {Phase diagram}. Each data point (${\color{red}\circ}$)
  corresponds to a set of (amplitude, frequency) parameter values for which BNE was observed 
  in our experiments, while the  points (${\color{blue}{\blacktriangle}}$) represent data sets where no bulk convection, and
therefore no BNE, was observed.  The solid line corresponds to $v=a (2 \pi
  f)=v_c$, while the dashed line corresponds to $\Gamma=a (2 \pi
  f)^2/g =1$. Bulk convection occurs only in the region where both the
  conditions $\Gamma > 1$ and $v > v_c$ are satisfied.}
\label{phase}
\end{center}
\end{figure}

Our understanding of the different parameter regimes for boundary
driven convection is summarized in the phase-diagram in
Fig.~\ref{phase}. We note that while $\Gamma >1 $ is necessary for
bulk convection, it is not a sufficient condition. As seen in the
phase diagram, there is a region of parameter values, with $\Gamma >
1$ and $v < v_c$, where no bulk convection occurs.  In experiments performed 
in this regime, we
observed some motion of the mustard seeds restricted entirely to the
the upper corners of the bed. Such a condition cannot give rise to BNE
for the intruder sizes studied here. We expect that for much larger
intruder sizes, other mechanisms may come into play which can drive
BNE \cite{duran94}.  We have verified (supplementary material) that the scaling of rise-time continues to hold for different bed-widths, wall roughness and 
for  choice of bed-particles. 
We also find  that on varying the bed-height, the scaling  holds for smaller heights (supplementary material) 
but breaks down for larger heights. 
The microscopic origins of  the parameters  $\alpha$, $A$ and $v_c$ are not clear as they depend on  the cell geometry which is easily
characterizable,  as well as   on details such as the wall,  
and bed-particle roughness, which are harder to characterize.

The scaling dependence of the velocity and the rise-time on the shaking 
velocity $a \omega$ can roughly be understood by analyzing the 
hydrodynamical equations that have been earlier used to describe vibrated granular 
systems.  In our case, the system is nearly incompressible in the bulk and hence 
the appropriate equations for the velocity field ${\bf u}(x,y)$ and the granular temperature $T(x,y)$, in the steady state, are \cite{grossman97,meerson03,alam10}
\begin{align}
{\bf u . \nabla u} &=-\frac{1}{\rho} {\bf \nabla} p 
+ \nu \nabla^2 {\bf u} + g\hat{\bf g} \\
{\bf u . \nabla} T &={\bf \nabla}.(\kappa {\bf \nabla} T) - I~, 
\end{align}
where $\rho$ is the ``granular fluid'' density, the pressure $p/\rho \propto   T/m$ and 
the viscosity and thermal conductivity   are given by 
$\nu \propto \kappa = (0.6 \ell +d)^2 \ell^{-1} \sqrt{T/m}$~, with $\ell$ and 
$d$ being the mean free path and particle diameter respectively.  
The energy dissipation is given by the term $I \propto \ell^{-1} T \sqrt{T/m}$.
With a rescaling of the variables ${\bf u} \to (a \omega) {\bf u},~\{p/\rho,T/m\} \to 
(a \omega)^2 \{ p/\rho,T/m \}$ and $\nabla \to b^{-1} \nabla$, where $b$ is some typical coarsening length scale, the above equations 
transform to
\begin{align}
{\bf u . \nabla u} &=-\frac{1}{\rho} {\bf \nabla} p 
+ \frac{\nu}{b}  \nabla^2 {\bf u} + \frac{b}{a}\frac{ \hat{\bf g}}{\Gamma} \label{hydsc1}\\
{\bf u . \nabla} T &=\frac{1}{b}{\bf \nabla}.(\kappa {\bf \nabla} T) - b I~. 
\end{align} 
For large $\Gamma$, the last term in Eq.~\eqref{hydsc1} can be neglected, and the scaled equations become independent of $a$ and $\omega$. This may explain the scaling behaviour of velocities with $a \omega$ seen in our experiments.

Finally, we anticipate that a similar
scaling should hold even in three-dimensional granular systems in some
parameter regime where granular convection is the dominant mechanism
driving the BNE.
We note that since the experiments in \cite{knight96} were performed at a fixed frequency, their result $\tau \sim (\Gamma - \Gamma_c)^{-2.5}$ is consistent with our  scaling form $\tau \sim (v-v_c)^{-\alpha}$ with $\alpha = 2.5$.

The authors thank P. Nandakishore and D. Saha for their help with the  experiments. We thank N. Menon and D. J. Durian for useful discussions.

\pagebreak
\renewcommand{\baselinestretch}{2}
\begin{widetext}

\section*{Supplementary material for  ``Scaling behavior in
  convection-driven Brazil-nut effect''}

\begin{figure}[h!]
\begin{center}
\includegraphics[width=6.5in]{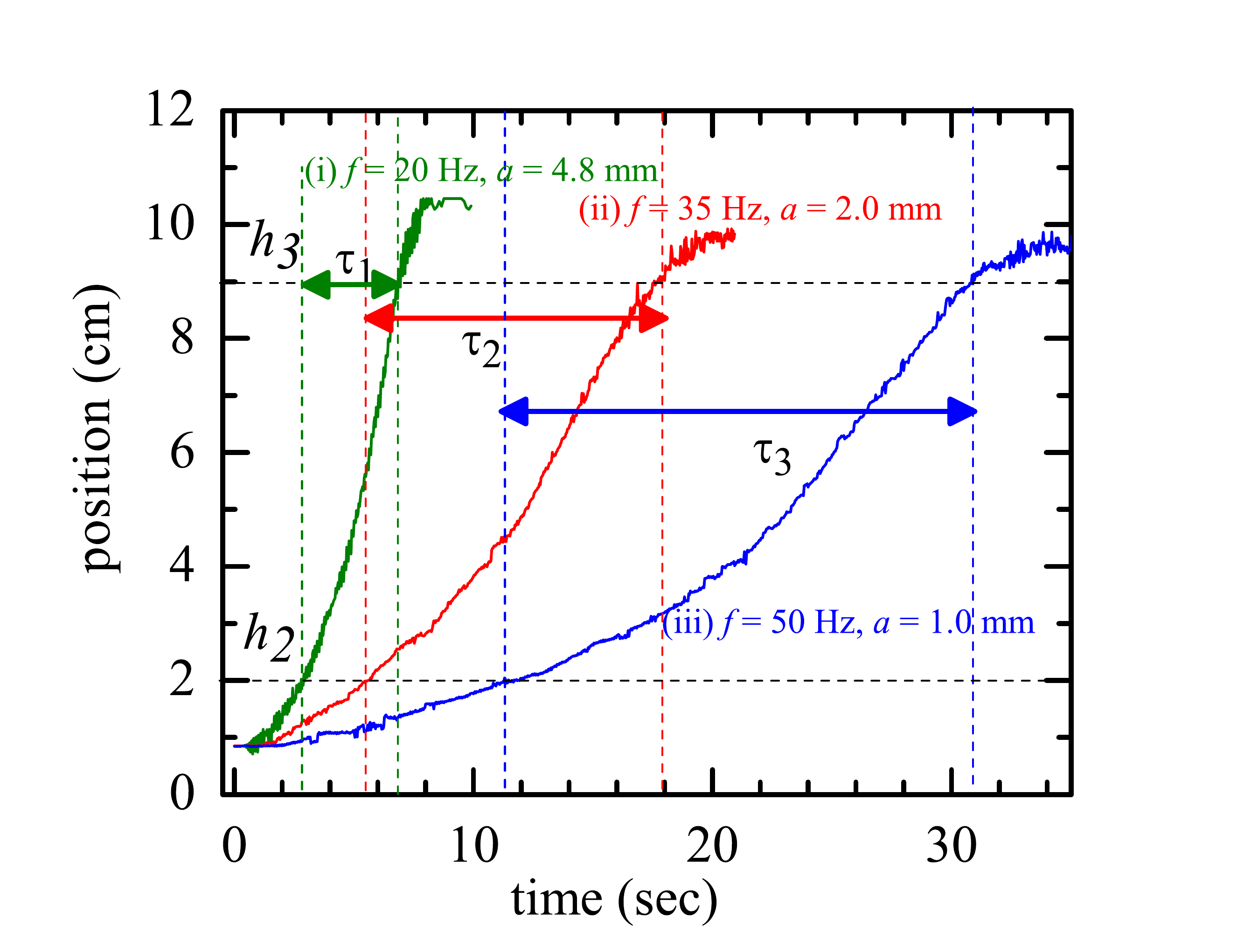}
\caption{{\bf Intruder position vs. time plot}. Position vs. time
  plots for an intruder of $\rho_r = 1.07$, for i) $f = 20$ Hz, $a =
  4.8$ mm (green), ii) $f = 35$ Hz, $a = 2.0$ mm (red) and $f = 50$
  Hz, $a = 1.2$ mm (blue). The bed-height was $10$ cm and width was $13$ cm and the sidewall roughness was $P-100$ sandpaper.
 The dashed horizontal lines (black)
  represent $h_{2}$ and $h_{3}$ (see Fig.~1{\bf a} of main paper).
  The rise-time of the particle is computed by estimating the time it
  takes for the intruder to rise from the position $h_{2}$ to $h_{3}$.
  The vertical dashed lines in green, red and blue enclose the time
  windows required by the intruder to rise between $h_{2}$ and $h_{3}$
  for the three shaking experiments (i), (ii) and (iii)
  respectively. }
\label{traj}
\end{center}
\end{figure}

\pagebreak

\begin{figure}
\begin{center}
\includegraphics[width=6.5in]{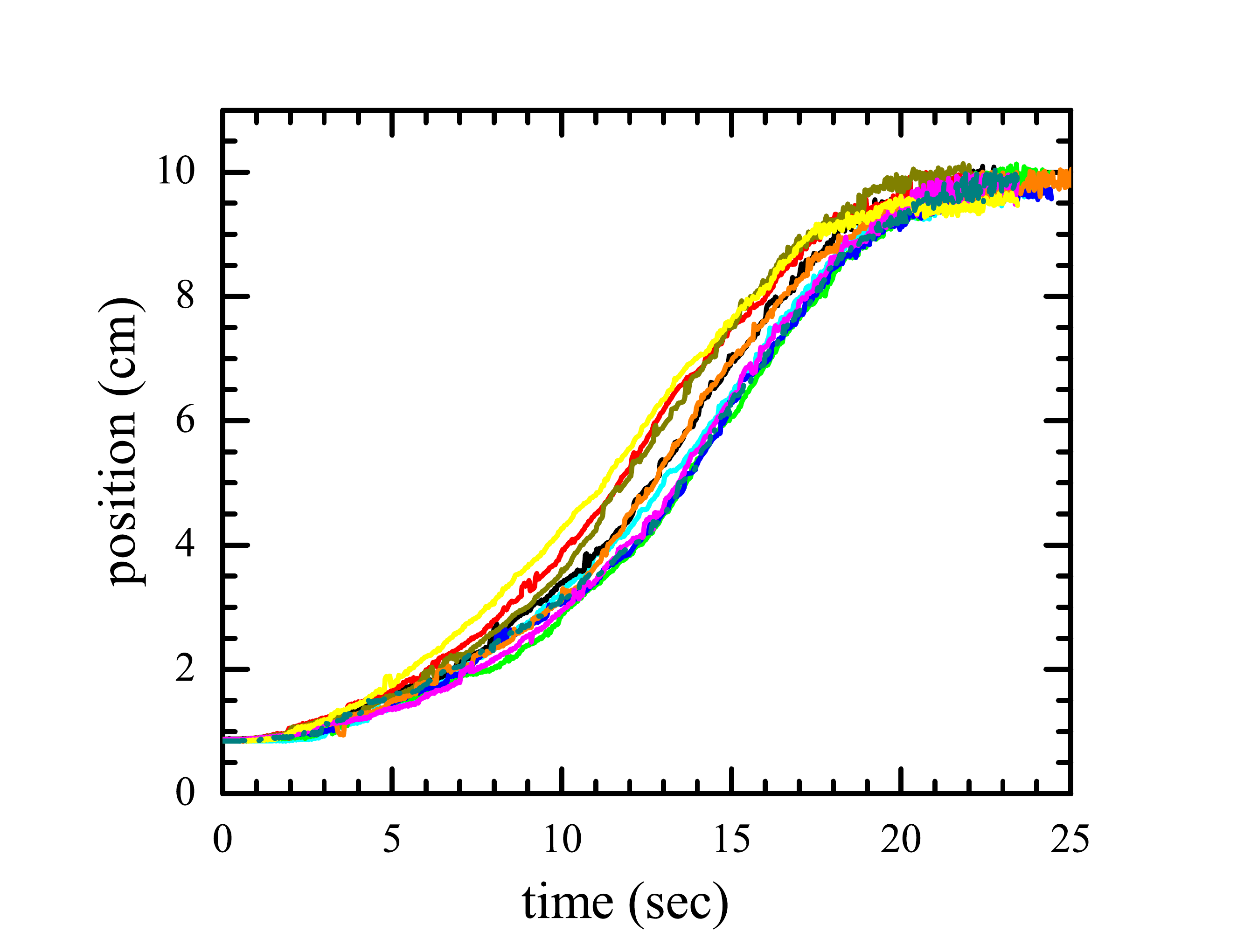}
\caption{ {\bf Different realizations of intruder
    trajectories}. Position vs. time plots for $10$ different trials
  of the rise of a $\rho_r=1.07$ intruder for $f=50$~Hz and $a=1.4$~mm.   
The bed-height was $10$~cm and width was $13$~cm and the 
sidewall was roughened with P-100 sandpaper.
 The trajectories show roughly quadratic dependence on time. The 
  different repeats of the experiment for fixed control parameters
  lead to slightly different trajectories and hence small fluctuations 
of rise-times over different realizations.  }
\label{trials}
\end{center}
\end{figure}

\begin{figure}
\begin{center}
\subfigure[~$f=20$~Hz, $a=3.7$~mm,
  $a\omega=46.5$~cm/sec]{\includegraphics[width=2.5in]{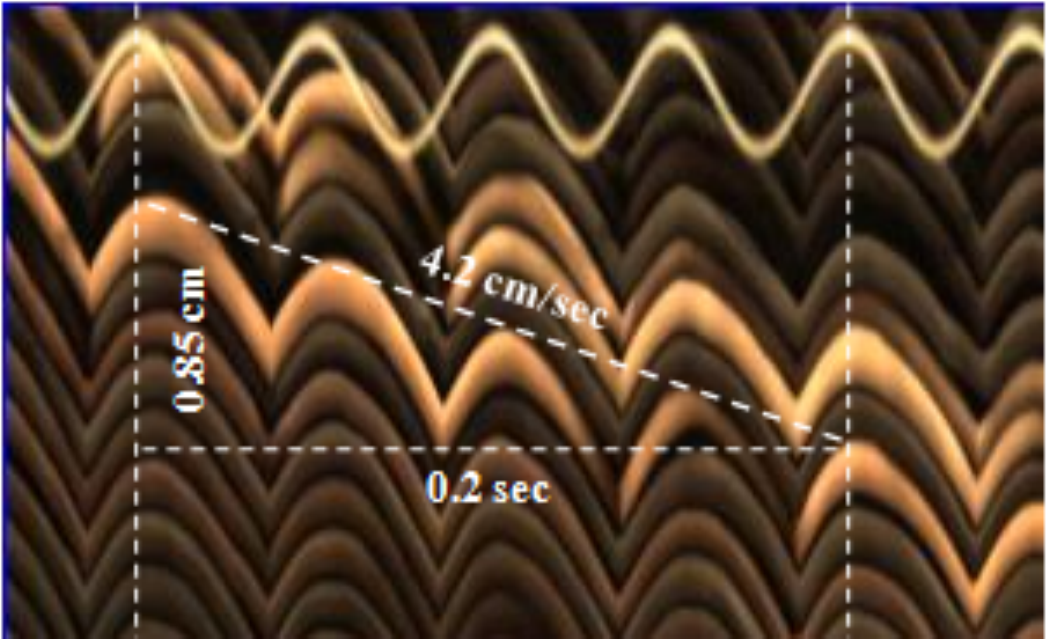}}
\hskip 1cm
\subfigure[~$f=35$~Hz, $a=2.1$~mm,
  $a\omega=46.2$~cm/sec]{\includegraphics[width=2.5in]{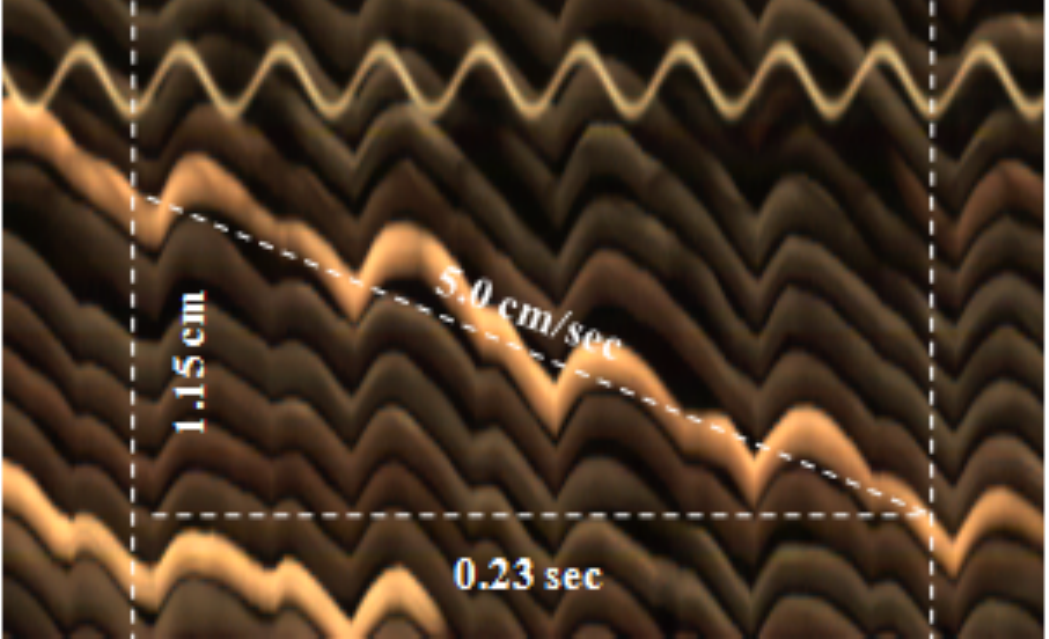}}
\subfigure[~$f=50$~Hz, $a=1.5$~mm,
  $a\omega=47.1$~cm/sec]{\includegraphics[width=2.5in]{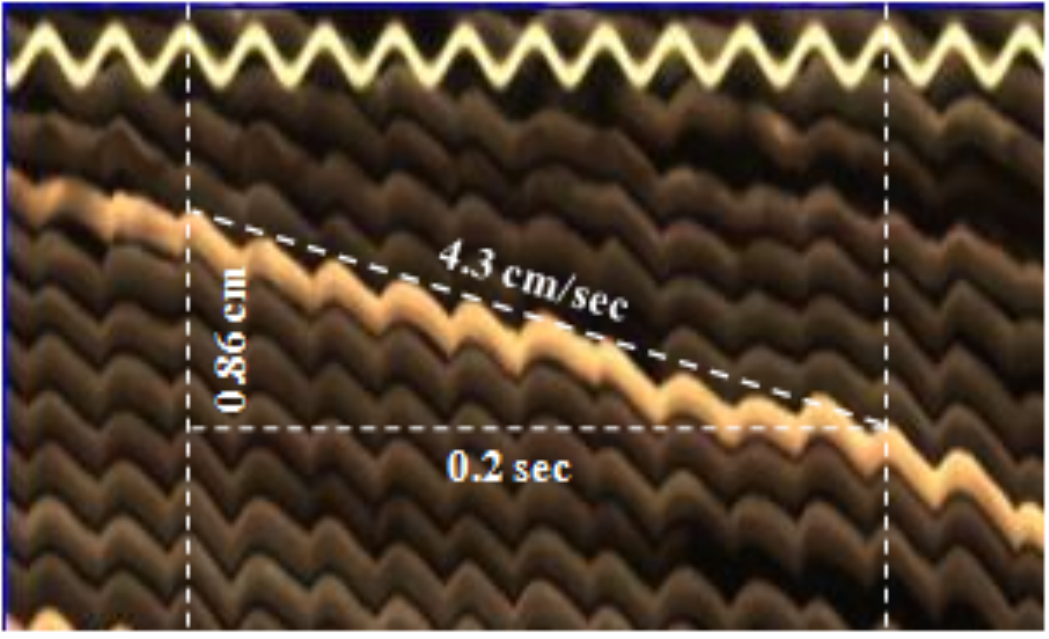}}
\caption{(Color online) {Tagged particle trajectories ($y$-vs-$t$) near
    wall}. Each bright, slanted yellow trajectory represents the
  time-evolution of the $y$ (vertical)-coordinate of a tagged particle
  (yellow mustard seed) adjacent to the side wall, for three different
  parameter sets.  
The slope of the yellow trajectory gives the velocity of the tagged
particles as (a) $4.2$~cm/sec, (b) $5.0$~cm/sec and (c) $4.3$~cm/sec
respectively. The white sinusoidal trace (horizontal) represents the
time-evolution of a reference white mark on the acrylic cell.}
\label{wallflow}
\end{center}
\end{figure}

\begin{figure}
\includegraphics[width=6.in]{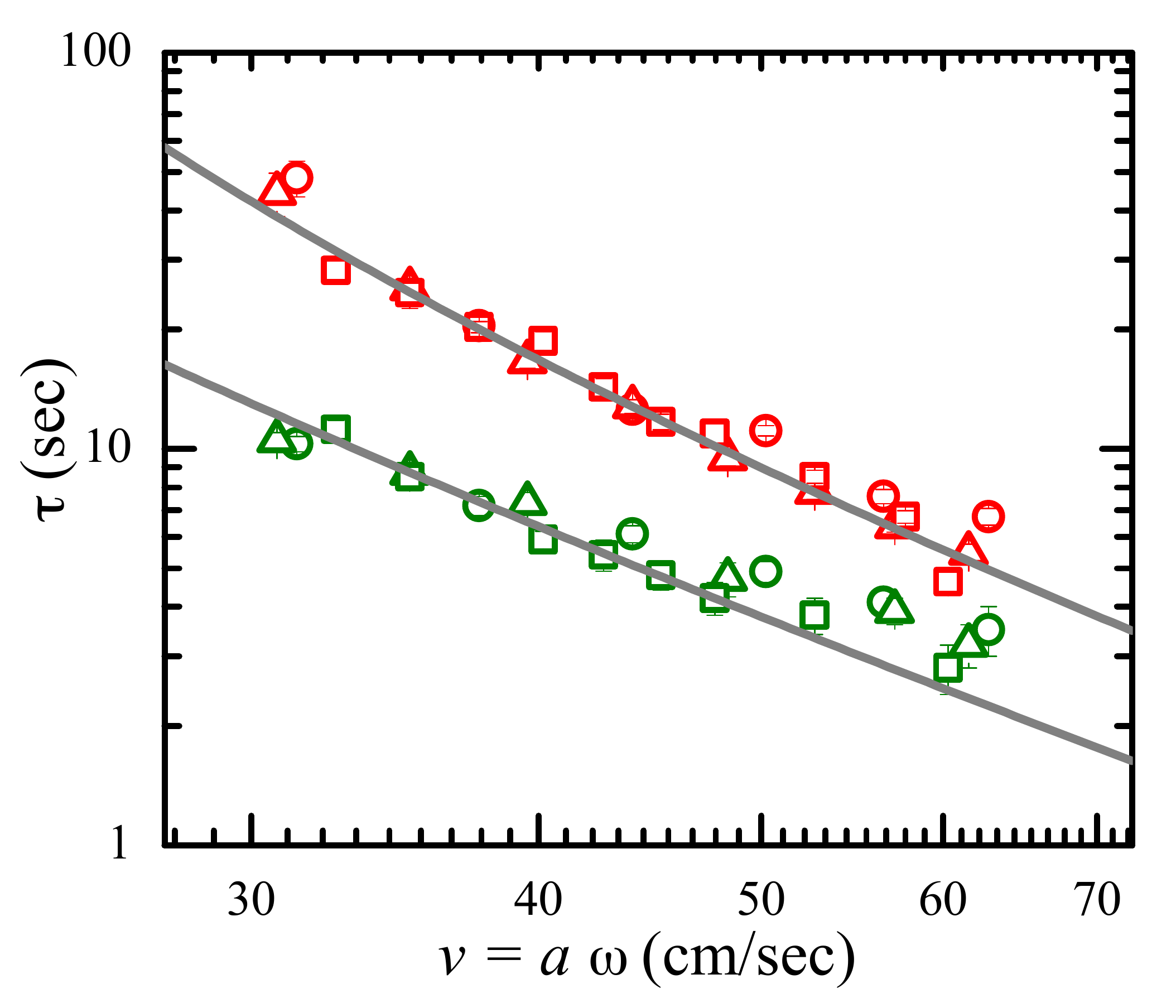}
\caption{{\bf Plots of the rise-time for different bed-heights}.  
     Plot of $\tau$ versus $v=a \omega$ for two different bed
  heights $h=8$~cm (green) and $h=10$~cm (red), 
  for three different frequencies $20$ Hz ($\square$), $35$ Hz
  ($\triangle$) and $50$ Hz ($\bigcirc$).  The bed-width was $13$ cm and the 
sidewall roughness was achieved with P-100
 sandpaper. In all 
  experiments, the relative density was $\rho_r=1.07$ and the intruder
  size was $1.7$ cm.  We see good collapse of data for both heights. 
The fits are to the scaling form $A(v-v_c)^{-2}$ with $A \approx 13000$ cm$^2$/sec, $v_c\approx 15$ cm/sec for the $10$ cm bed and $A \approx 7000$ cm$^2$/sec, $v_c\approx 7$ cm/sec for the $8$ cm bed.}  
\label{bedheights}
\end{figure}

\begin{figure}
\includegraphics[width=6.in]{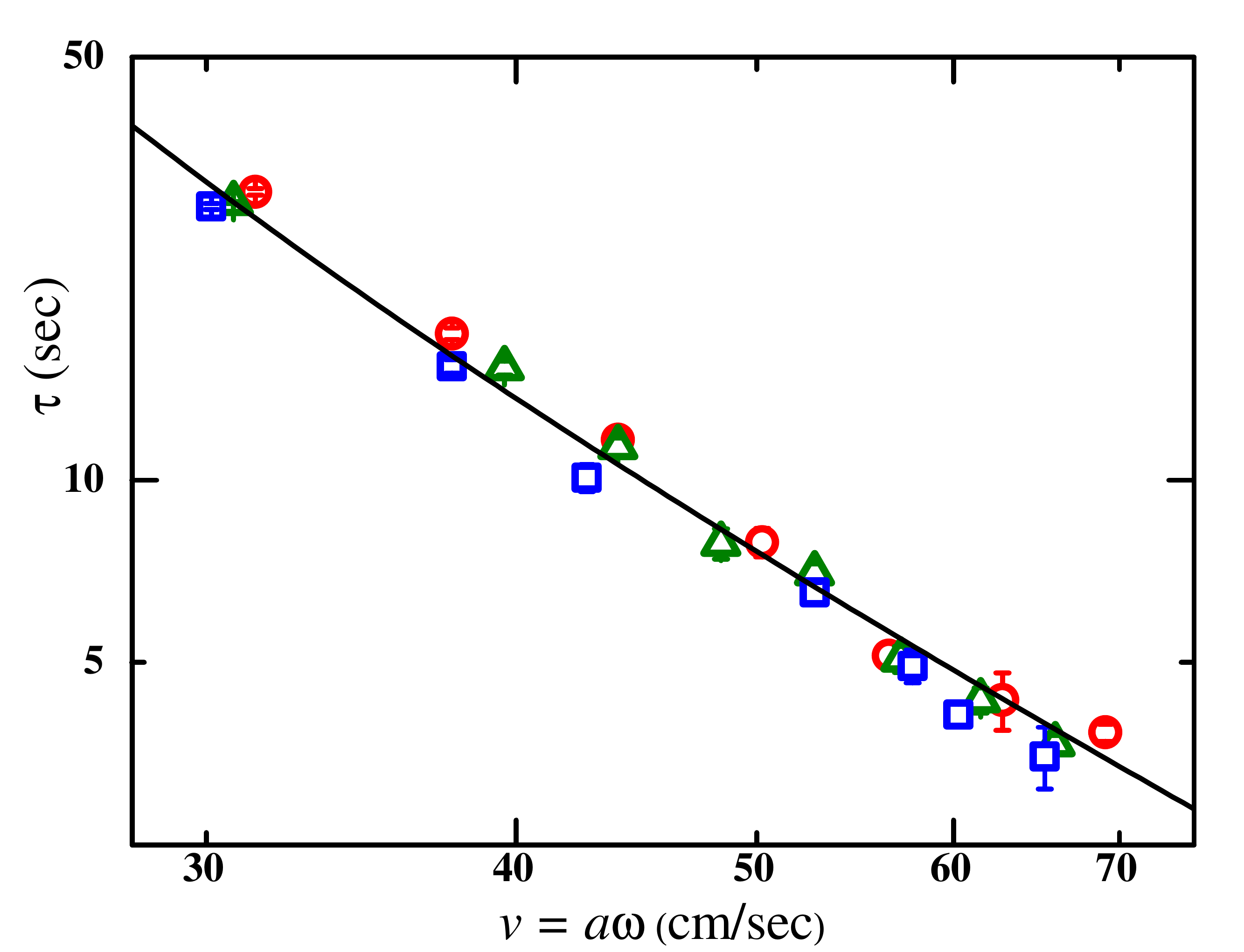}
\caption{{\bf Plots of the rise-time for a different choice of bed-particles.}
Here the bed particles were fenugreek seeds that are larger, have
different shapes and are characterized by higher size polydispersity than
the mustard seeds.
The bed-height was $h=10$ cm, bed-width $13$ cm, sandpaper  roughness $P-100$ and the data is   for three different frequencies $20$ Hz (${\color[rgb]{0,0.7,0}\bm{\square}}$), $35$ Hz
  (${\color[rgb]{0,0,0.9}\bm{\triangle}}$) and $50$ Hz (${\color[rgb]{0.9,0,0}\bm{\bigcirc}}$). The intruder had a diameter $1.7$~cm, thickness $5$~mm and density $1.07$~g/cc. }
\end{figure}

\begin{figure}
\includegraphics[width=6.in]{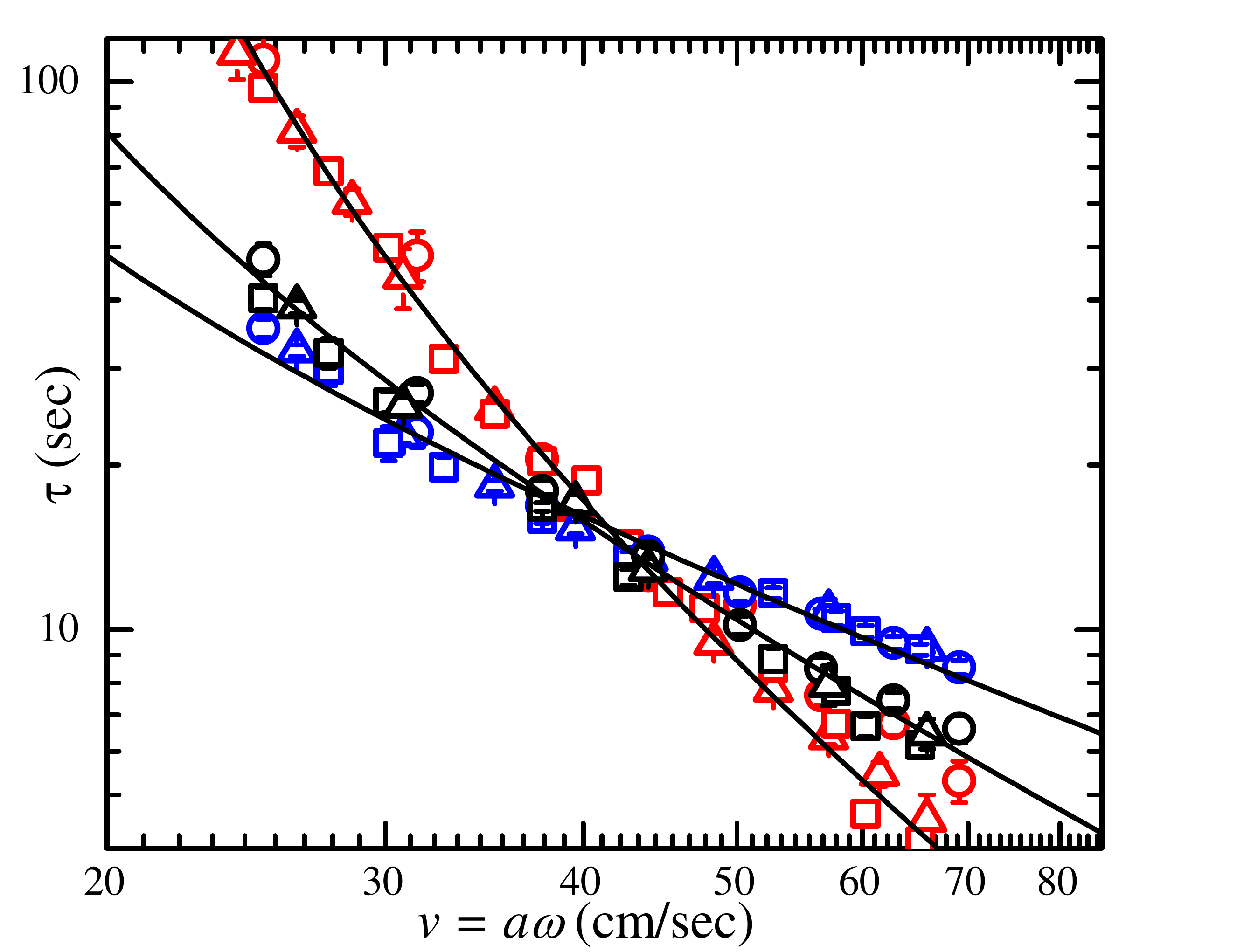}
\caption{{\bf Plots of the rise-times for three different  bed-widths}. The data sets correspond to the  cell widths $13$ cm (red), $15.5$ cm (black) and $18$ cm (blue), for three different frequencies $20$ Hz ($\square$), $35$ Hz
  ($\triangle$) and $50$ Hz ($\bigcirc$). The bed-particles were mustard seeds, cell-height was $10$ cm and sidewall sandpaper roughness $P-100$.  The intruder had a diameter $1.7$~cm, thickness $5$~mm and density $1.07$~g/cc.  The solid lines correspond to the 
scaling form $(v-v_c)^{-\alpha}$ with $\alpha=2.0$, $1.4$ and $1.0$ respectively.
}
\end{figure}

\begin{figure}
\includegraphics[width=6.in]{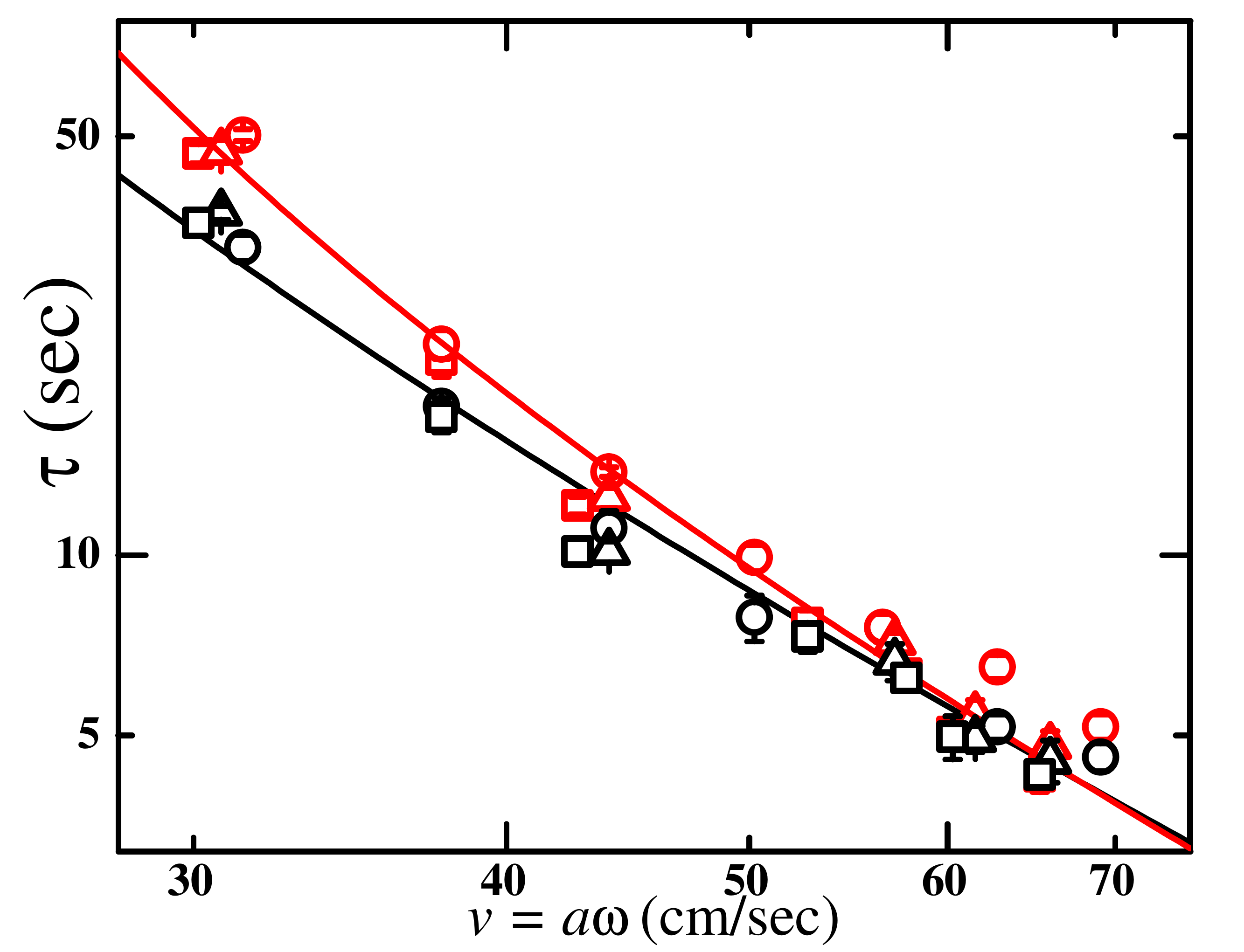}
\caption{{\bf Plots of the rise-times for two different side-wall roughnesses}. 
Here we show the rise-time data for two different wall roughness using 
sandpaper with ISO grit designations $P-100$ (red) and $P-60$ (black).  The bed-particles were mustard seeds, cell-height was $10$ cm and bed-width was $13$ cm. The intruder had a diameter $1.7$~cm, thickness $5$~mm and density $1.07$~g/cc. The 
solid lines correspond to 
the scaling form    $(v-v_c)^{-2}$ with $v_c=15$ cm/sec and $10$ cm/sec 
respectively.}

\end{figure}

\end{widetext}


\begin{thebibliography}{99}

\bibitem{rosato87} A. Rosato, K. J. Strandburg, F. Prinz, and R. H. Swendsen,
  Phys. Rev. Lett. {\bf 58}, 1038 (1987).

\bibitem{JNB96} H.M. Jaeger, S.R. Nagel, and R.P. Behringer, 
   Rev. Mod. Phys. {\bf 68}, 1259 (1996).

\bibitem{mobius01} M. E. M\"obius, B.E. Lauderdale, S.R. Nagel, and
  H.M. Jaeger,  
 Nature (London) {\bf 414}, 270 (2001).


\bibitem{kudrolli04} A. Kudrolli, 
Rep. Prog. Phys. {\bf 67}, 209 (2004).   

\bibitem{duran94} J. Duran, T. Mazozi, E. Cl\'ement, and J. Rajchenbach, 
Phys. Rev. E {\bf 50}, 5138 (1994).

\bibitem{liffman01} K. Liffman {\emph et al.},  
Granular Matter {\bf  3}, 205 (2001). 

\bibitem{knight96}
 J. B. Knight, E. E. Ehrichs, V. Y.  Kuperman, J. K. Flint, H. M. Jaeger and S. R. Nagel, 
 Phys. Rev. E {\bf 54}, 5726 (1996).


\bibitem{vanel97} L. Vanel, A. D. Rosato, and R. N. Dave, 
Phys. Rev. Lett. {\bf 78}, 1255 (1997).   

\bibitem{cooke96} W. Cooke, S. Warr, J. M. Huntley, and R. C. Ball, 
Phys. Rev. E {\bf 53}, 2812 (1996).


\bibitem{schroter06} M. Schr\"oter, S. Ulrich, J. Kreft, J. B. Swift, and
  H. L. Swinney, 
Phys. Rev. E {\bf 74}, 011307 (2006).


\bibitem{knight93} J. B. Knight, H. M. Jaeger and S. R. Nagel,  
  Phys. Rev. Lett. {\bf 70}, 3728 (1993).
  


\bibitem{knight97} J. B. Knight, 
Phys. Rev. E {\bf 55}, 6016 (1997).
 


\bibitem{BM95} M. Bourzutschky and J. Miller,  
Phys. Rev. Lett. {\bf 74},  2216 (1995).
 
\bibitem{ehrich95}  E. E. Ehrichs, H. M. Jaeger,  G. S. Karczmar,
  J. B. Knight,    V. Y. Kuperman and  S. R. Nagel, 
 Science  {\bf 267},  1632
  (1995). 





\bibitem{eshuis05} P. Eshuis, K. van der Weele, D. van der Meer, and
  D. Lohse,  
Phys. Rev. Lett. {\bf 95}, 258001 (2005).
  

\bibitem{eshuis07} P. Eshuis, K. van der Weele, D. van der Meer, R. Bos and D. Lohse, 
 Phys. Fluids {\bf 19}, 123301 (2007).


\bibitem{alam10} P. Eshuis, D. van der Meer, M. Alam, K. van der Weele and D. Lohse, 
Phys. Rev. Lett. {\bf 104}, 038001
  (2010). 

\bibitem{StAndrews} http://grahammilne.com/tracker.htm


\bibitem{mobius04} M. E. M\"obius, X. Cheng, G. S. Karczmar, S. R. Nagel, and
  H. M. Jaeger,  Phys. Rev. Lett. {\bf 93}, 198001 (2004).   


\bibitem{mobius05} M. E. M\"obius, X. Cheng, P. Eshuis, G. S. Karczmar,
  S. R. Nagel, and   H. M. Jaeger,  Phys. Rev. E {\bf72}, 011304 (2005).

\bibitem{dijksman09} J. A. Dijksman and M. van Hecke, Euro. Phys. Lett. {\bf 88}, 44001 (2009).

\bibitem{grossman97} E. L. Grossman, T. Zhou, and E. Ben-Naim, Phys. Rev. E {\bf 55}, 4200 (1997).


\bibitem{meerson03} B. Meerson, T. P\"oschel, and Y. Bromberg, Phys. Rev. Lett. {\bf 91}, 024301 (2003).








\end{thebibliography}
\end{document}